# ON THE OMNIPRESENT BACKGROUND GAMMA RADIATION OF THE CONTINUOUS SPECTRUM

R, Banjanac, D. Maletić, D. Joković, N. Veselinović,
A. Dragić, V. Udovičić, I. Aničin

*Institute of Physics, University of Belgrade, Pregrevica 118, 11080 Belgrade, Serbia*
*e-mail: banjanac@ipb.ac.rs*

**Abstract:** The background spectrum of a germanium detector, shielded from the radiations arriving from the lower and open for the radiations arriving from the upper hemisphere, is studied by means of absorption measurements, both in a ground level and in an underground laboratory. It is established that the continuous portion of this background spectrum is mostly due to the radiations that arrive from the upper hemisphere of the continuous spectrum similar to the instrumental one. The intensity of this radiation of an average energy of about 120 keV is estimated to about 8000 photons/m$^2$s 2π srad in a ground level laboratory, and to about 5000 photons/m$^2$s 2π srad at the depth of 25 m.w.e. Rough estimates of the dose that it contributes to the skin are about 1.5 nSv/h and 1 nSv/h respectively. Simulations by GEANT4 and CORSIKA demonstrate that this radiation is both of cosmic and terrestrial origin, mixed in proportion that still has to be determined.

1. **Introduction**

Main contributors to background spectra of unshielded germanium detectors are the gamma radiations of discrete spectrum that originate from naturally occuring radioactive isotopes dispersed in the environment and the materials that surround the detector, and the radiations whose origin can be traced to cosmic rays. Gamma radiations of discrete energies produce the line spectrum but are also partially responsible for the continuum, composed of the Compton distributions of discrete energies that escape total detection. Due to the intrinsically high peak-to-Compton ratio, this continuum is in germanium detectors generally much lower than the lines that produce it. The bremsstrahlung due to the presence of Pb-210 in lead castles may be contributing to the continuum. Cosmic-ray muons produce the continuous spectrum of energy losses that, for all detector sizes but for the thinnest ones, peaks at high energies, well beyond the region where the spectrum is usually of interest. The muon secondaries, however, contain significant quantity of low-energy radiations that contribute to the continuum in its portion relevant to spectroscopy. The soft, electromagnetic component of cosmic rays by its scattered and degraded radiations also contributes to the continuous part of the background spectrum, mostly at lower energies, within the region of interest to practical spectroscopy. Neutrons, mostly of cosmic-ray origin, contribute the continuous spectrum of recoils that diverges at lowest energies, though usually of very low intensity. The only spectral line that is attributed to cosmic rays is the annihilation line. All this results in the instrumental background spectrum characteristic of the detector size, shape and the dead layers. The prominent feature common to all instrumental background spectra, however, is that the greatest part of their spectral intensity lies in the low-energy continuum that, depending chiefly on the detector size, peaks at around 100 keV. In this work we study by means of absorption measurements the background radiations that arrive from the upper hemisphere, which may be considered responsible for the greatest part of this continuum, with the aim of determing its intensity and origin.

2. **The experiment**

The measurements were performed with a 35% efficiency coaxial type radio-pure HPGe detector mounted in a 1 mm thick magnesium housing, shielded by the lower half of a heavy lead castle from the radiations coming from the lower hemisphere and completely open to those arriving from the upper hemisphere. The same setup was used in both the ground level and in the underground laboratory (at 25 m.w.e.). The laboratories are described in some detail in [1]. A set of measurements is performed with lead absorbers of increasing thickness positioned so as to block the way to the radiations coming from above. The background spectra from such measurements are presented in Fig.1. Absorber thicknesses are marked in the figures, which are presented in two different scales, to emphasize the general change of spectra upon absorption, in the figures on the left, and the details around the X-rays of lead, in the figures on the right.

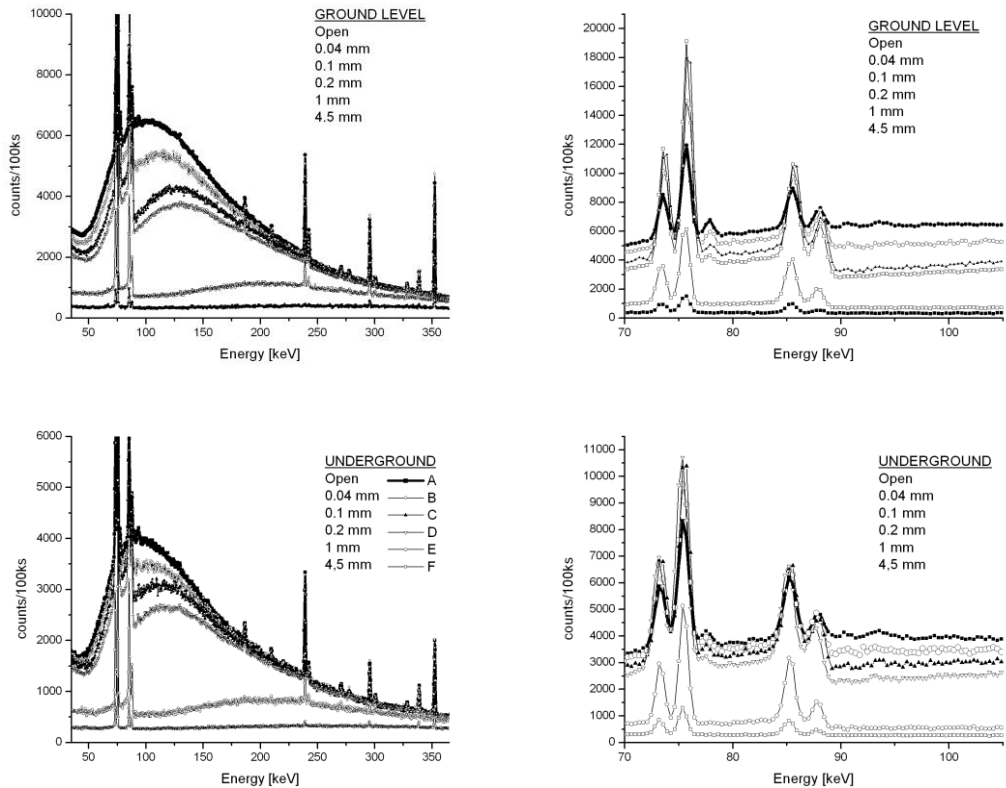

**Fig.1.** Experimental low-energy portions of background spectra of a HPGe detector completely shielded from the radiations coming from the lower hemisphere, with a set of lead absorbers of different thickness positioned so as to intercept the radiations arriving at the detector from the upper hemisphere. Upper figures are with the detector located in a ground level laboratory while the lower ones are taken in the underground laboratory at the depth of 25 m.w.e. All spectra are normalized to the measurement time of 100 ks. Visual inspection already suggests that the first half-thickness at the energy that carries maximum intensity is different on the surface and underground. This suggests that, in spite of the apparently similar character of the spectra, their composition is different on the surface and underground.

3. **The results**

Visual inspection of the absorption spectra presented in Fig. 1 leads to a number of interesting qualitative conclusions: **1.** The spectra taken on the ground level and in the underground exhibit great similarity, the integral intensity of the continuum in the underground being about 1.75 times smaller. At the same time the intensity of cosmic-ray muons in the underground is about 3.5 times smaller [2]. **2.** The energy which carries maximum intensity in the continuum increases with absorber thickness, what is typical of continuous spectra, and is known as the "hardening of the spectrum". **3.** The discontinuity in the absorption spectra on the energy of $K_\beta$ X-rays of lead (absorption edge) reflects the fact that the instrumental continuous spectrum is mostly due to the radiations of the same continuous spectrum, and not due to incomplete detection of radiations of higher discrete energies. **4.** Initial increase of the intensity of fluorescent X-rays of lead with absorber thickness again witnesses that the incoming radiation is absorbed by the photoelectric effect, and that the real spectrum of this radiation is similar to the instrumental one, at least to the energies of about 200 keV, where the photoelectric effect in lead dominates over the Compton effect. **5.** Some apparent differences in absorpiton character of the spectra taken on the ground level and in the underground are to be expected on account of necessarily different composition of the radiations and their different angular distributions at the two locations.

These qualitative conclusions are supported by quantitative analyses of absorption curves at different energies of the continuum. As an illustration, Fig.2 presents the absorption curves for the count in the channel in the continuum that correspons to energy of 89 keV, close to the K-absorption edge in lead.

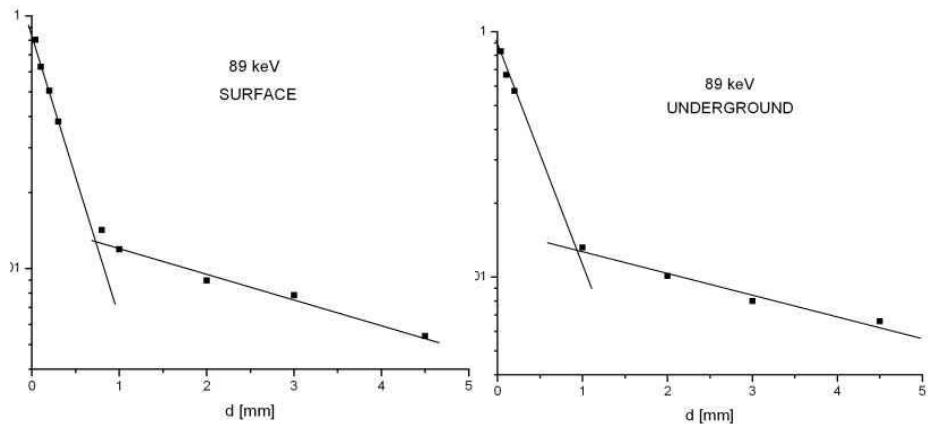

**Fig.2.** Absorption curves for the count in the continuum that corresponds to energy of 89 keV, in the ground-level laboratory (left) and in the underground laboratory at 25 m.w.e. (right). Two distinct components are seen; the first corresponds rather well to this same energy, while the other, much more penetrating, approximately corresponds to an average energy of about 500 keV.

The two well-defined components of very different absorption properties are found. On the surface, the less penetrating one by its absorption coefficient corresponds within the errors to the energy close to 90 keV, while the same component in the underground appears of slughtly different absorption properties, due to necessarily different composition of the radiations and their different angular distributions. The much less intense and much more penetrating component both on the surface and in the underground roughly corresponds to an energy of about 500 keV. The first component thus represents the radiation of the same energy at which it appears in the spectrum, which belongs to the continuum, while the second one represents the sum of Compton distributions of all radiations of higher energies that escape full detection, which thence manifest absorption properties of the radiation of an average energy that in our case appears to be around 500 keV. Since the low-energy component is practically fully absorbed by 1 mm of lead, subtracting the spectrum that corresponds to the absorber of that thickness from the spectrum of the open detector approximately leaves only the spectrum of the radiations of the genuinely continuous spectrum, as seen by a given detector. These spectra are presented in Fig.3.

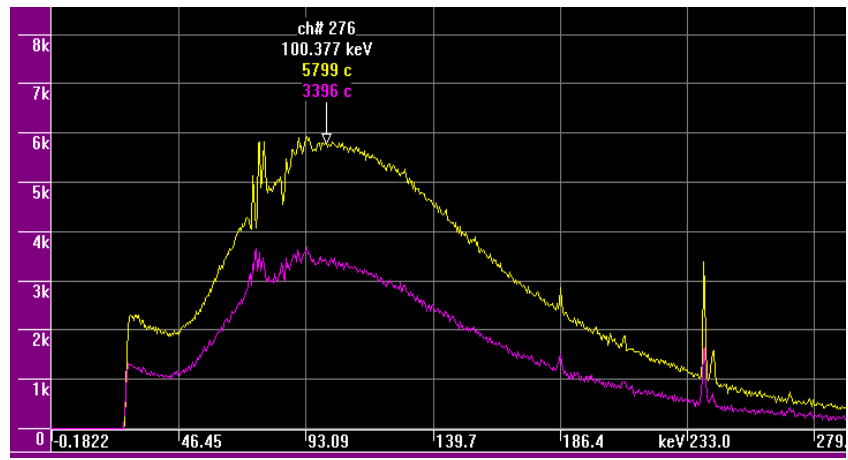

**Fig.3.** Approximate shapes of instrumental spectra of background radiations of the continuous spectrum arriving at a HPGe detector open towards the upper hemisphere in a ground-level laboratory (upper spectrum) and in an underground laboratory at 25 m.w.e. (lower spectrum). The lines are residuals due to effects that are unessential here. Integral count rates in these spectra are 21 cps and 12 cps respectively.

The integrals of these spectra, very roughly corrected for absorption in the detector housing and the detector dead layers, yield for the fluxes of these radiations that arrive from the upper hemisphere with an average energy slightly over 100 keV the values of about 8000 photons/$m^2$.s.$2\pi$ srad on the surface, and about 5000 photons/$m^2$.s.$2\pi$ srad in the underground. Simplifying assumption that this radiation is by its ionization properties similar to the radiation of Co-57 yields an estimate of the dose that it delivers to the skin of about 1.5 nSv/h on the surface and about 1 nSv/h in the underground. An important property of these spectra is that the maximum of intensity at around 100 keV, as well as the dip of intensity at an energy of about 40 keV, is an essential property of the true spectrum of the incoming radiations, and is only partly due to the drop of detection efficiency at these energies. It also seems that the steep increase of intensity below the dip is an intrinsic property of all these spectra. We could not reach this region but there is ample evidence in background spectra taken at other places (e.g. see ref. 3) that this is also their ubiquitous property.

All these conclusions are qualitatively corroborated by the detailed simulations of the experimental situations that result in these spectra using the GEANT4 and CORSIKA simulation packages. Two contributions to these spectra were considered. The first is the contribution of environmental natural radioactivity via the scattering of discrete energy gamma rays off the air, the walls, and the ceiling, that thus produces the so called skyshine radiation, which is known to be of spectral shape similar to that of our Fig.3 [4]. For this simulation the as realistic as possible distribution of radioactivities was assumed in the environment and the configuration of the detector was taken into account in detail. The result of this simulation of very low efficiency is, for the setup in the underground laboratory, presented in Fig.4 (equivalent of 4000 hours of single cpu time went into the production of this figure).

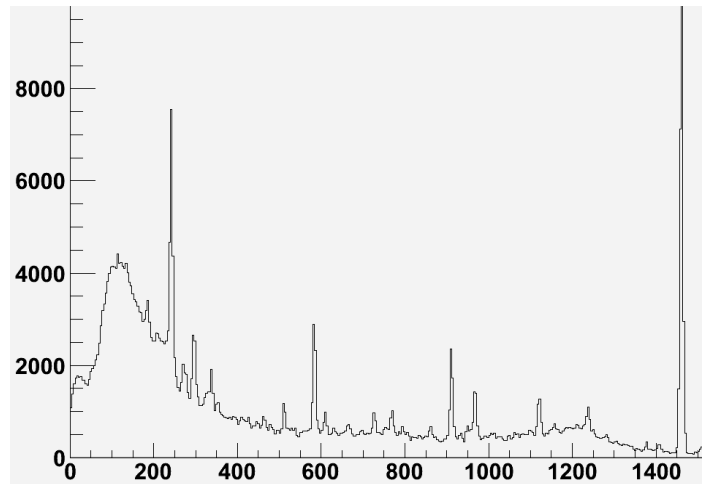

**Fig.4.** The "skyshine" spectrum due to environmental radioactivity in the underground laboratory according to the rather realistic simulation by GEANT4 of the experimental situation that produced the spectrum of Fig.1, lower portion. Similarity of the continuous parts of the two spectra is obvious. Absolute intensities are however difficult to compare.

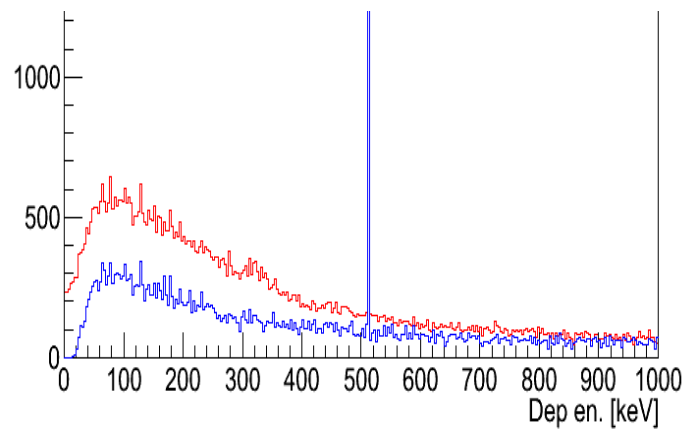

**Fig.5.** Total contribution of all radiations of cosmic-ray origin to the low-energy part of the background spectrum of the detector setup in the underground (higher spectrum), as simulated by CORSIKA down to the plane at the surface of the Earth, and then taken over by GEANT4 all the way to the detector spectrum. Basic features of the experimental spectrum of Fig.1 are again reproduced, but it is difficult to extract relative contributions of the skyshine radiation and cosmic rays to this spectrum.

Finally, Fig.5 presents the results of the simulation by CORSIKA+GEANT4 of all the contributions due to cosmic rays to the background spectrum in the underground. The shape of the spectrum again corresponds well to the experimental one, but it is equally difficult to establish its absolute intensity. One possibility would be to normalize the simulated spectrum of muon energy losses to the experimental peak, which would be found at about 40 MeV, but it is again inaccessible with our detector.

4. **Conclusions**

We have established that the low-energy continuous part of background spectra of open germanium detectors that peak around 100 keV can in a good approximation be considered as being due to the radiations of the similar continuous spectrum that arrive from the upper hemisphere, and that it is therefore in the greatest part absorbed by 1 mm of lead. This is valid both in a ground level and in the underground laboratory at 25 m.w.e. The origin of this radiation is of both terrestrial and cosmic ray origin, in proportion that still remains to be determined.

This work is supported by the Ministry for Education, Science and Technological Development of Serbia under the Project OI 171002.

**References**

1. Dragić A *et al*, The new setup in the Belgrade low-level and cosmic-ray laboratory, *Nucl. Tech and Rad. Prot.* **26**(2011)181-192, http://ntrp.vinca.rs/2011_3/Dragic2011_3.html, and arXiv:1203.4607
2. Dragić A *et al*, Measurement of the cosmic ray muon flux in the Belgrade ground level and underground laboratories, Nucl. Instr, and Meth. In Phys. Res. **A591**(2008)470-475
3. Budjaš D et al, Highly sensitive gamma-spectrometers of GERDA for material screening: part 2, Proc. XIV Baksan School 2007, *INR RAS, Moscow 2008. ISBN 978-5-94274-055-9*, pp. 233-238 and arXiv:0812.0768
4. Swarup J, Photon spectra of gamma rays backscattered by infinite air II, Skyshine, *Nucl. Instr. and Meth*. **172**(1980)559-566